\documentclass[preprint,showpacs,preprintnumbers,amsmath,amssymb]{revtex4}

\usepackage{graphicx}% Include figure files
 \usepackage{dcolumn}% Align table columns on decimal point
\usepackage{bm}% bold math

\begin{document}

%\preprint{APS/123-QED}

\title{Evaluating residues and integrals through \\
Negative Dimensional Integration Method (NDIM)}% Force line breaks with \\

\author{Alfredo Takashi Suzuki}
 \altaffiliation[Permanent address: ]{Instituto de F\'{\i}sica Te\'orica,\\
 Universidade Estadual Paulista, \\Rua Pamplona, 145 - 01405-900 Sao Paulo, SP}%Lines break automatically or can be forced with \\
\affiliation{Department of Physics
\\
North Carolina State University\\
Raleigh, NC 27695-8202}

%\affiliation{

\date{\today}% It is always \today, today,

\begin{abstract}
The standard way of evaluating residues and some real integrals
through the residue theorem (Cauchy's theorem) is well-known and
widely applied in many branches of Physics. Herein we present an
alternative technique based on the negative dimensional integration
method (NDIM) originally developed to handle Feynman integrals. The
advantage of this new technique is that we need only to apply Gaussian
integration and solve systems of linear algebraic equations, with no
need to determine the poles themselves or their residues, as well as
obtaining a whole class of results for differing orders of poles
simultaneously.
\end{abstract}

\pacs{Valid PACS appear here}% PACS, the Physics and Astronomy
\maketitle

\section{Introduction}

In a textbook on complex variables we may find real definite integrals
of the type
\begin{equation}
\label{I1I2}
I_1=\int_0^\infty \! \frac{dx}{x^2+1} \qquad \mbox {or} \qquad
I_2=\int_0^\infty \! \frac{dx \; x^2}{(x^2+1)(x^2+4)}
\end{equation}
to be evaluated using the contour integration in the complex plane,
making use of the Cauchy's residue theorem. This is, of course, a
simple exercise in complex analysis, and residue theorem is a powerful
tool to handle such integrals. However, if we were actually to
evaluate them we would do it separately, each integral in its turn,
with its own residues of poles summed over to get the final
answer. Not so if we use the NDIM technique, as we shall shortly
see. In NDIM we can integrate both integrals at the same time, and
more, without having to do it one by one separately;
we simply need to evaluate the general integral 
\begin{equation}
\label{I1I2a}
I_g^{(m,n,p)}=\int_0^\infty \! \frac{dx \; (x^2)^m}{(x^2+1)^n(x^2+4)^p}
\end{equation}
and then work out the result for a particular set of numbers $m,n,p=0,1,2,3,...$, of which $I_1$ and $I_2$ are specific examples.

The concept of negative dimensional integration can be best seen in
the following $D$-dimensional Gaussian integration:
\begin{equation}
\label{gaussian}
G=\int \! d^D\!x\:{\rm e}^{-\alpha x^2} = \left(\sqrt
{\frac{\pi}{\alpha}}\right)^D
\end{equation}

On the other hand, expanding in power series the exponential function
in the integrand of (\ref{gaussian}) above, we have
\begin{eqnarray}
\label{powerseries}
G&=&\int \! d^D \!x \:{\rm e}^{-\alpha x^2}\nonumber \\ &=&\int \! d^D \!x
 \sum_{n=0}^\infty (-)^n \frac{\alpha ^n}{n!}(x^2)^n\nonumber \\
 &=&\sum_{n=0}^\infty (-1)^n \frac{\alpha ^n}{n!}\int \! d^D \!x (x^2)^n
\end{eqnarray}

Now, comparing (\ref{gaussian}) and (\ref{powerseries}), we conclude
that for the equality to hold we must have
\begin{equation}
\label{main}
{\cal I}\equiv \int \! d^D\!x (x^2)^n = (-1)^{-n} n!\;(\sqrt{\pi})^{D}\;\delta_{n+\frac{D}{2},0}
\end{equation}
and since, by construction $n\geq 0$, one is led to assume that $D$ is
a negative valued dimension.
 
This non-trivial result for a $D$-dimensional integral with pure
quadratic integrand elevated to a given power was first pointed out by
Ricotta and Halliday in \cite{ricotta}, which differed somewhat from
all the previous considerations, starting from dimensional
regularization scheme developed by 't Hooft and Veltman \cite{thooft},
where all such integrals were set to zero straighforwardly. Since
Feynman loop diagrams lead to Feynman integrals in $D$-dimensions in
the dimensional regularization scheme, integrals evaluated with the
help of (\ref{main}) in negative dimensions, must be brought back by
analytic continuation to the realm of positive dimensions $D$. Since
the concept of negative dimensional integral arises as a consequence
of the positiveness of the polynomial powers in the integrands, the
mentioned analytic continuation is achieved by employing a property of
the Pocchammer's symbols bearing these parameters labeling powers of
the polynomial expressions in the integrands, namely
\cite{labtesting}
\begin{equation}
\label{Poch}
(a)_{-k} = \frac{(-)^k}{(1-a)_k}
\end{equation}
where
\[
(a)_k = \frac{\Gamma (a+k)}{\Gamma (a)}
\]

Of course, here we strict ourselves to one-dimensional integrals such
as (\ref{I1I2a}), so that (\ref{main}) becomes
\begin{equation}
\label{main1d}
{\cal I}_{D=1}\equiv \int\! dx (x^2)^n = (-1)^{-n}
n!\;\sqrt{\pi}\;\delta_{n+\frac{1}{2},0}
\end{equation}
and all the NDIM formulation developed for $D$-dimensional Feynman integrals can be applied to evaluate one-dimensional integrals of the type $I_1$ and $I_2$.

\section{Review of Cauchy's residue theorem application}
Consider, for practical example, the integral
\begin{equation}
\label{I1}
I_1=\int_0^\infty \!
\frac{dx}{x^2+1}=\frac{1}{2}\int_{-\infty}^{+\infty}\!\frac{dx}{x^2+1}
\end{equation}
 
The second integral on the RHS of the above represents an integration
along the real axis of the function
\begin{equation}
f(z)=\frac{1}{z^2+1}=\frac{1}{(z+i)(z-i)}
\end{equation}
which is a function of complex variables with simple poles at $z=\pm
i$.

Let $C_R$ be the semi-circle with radius $|z|=R$ with $R>1$ as shown
in Fig. 1 below

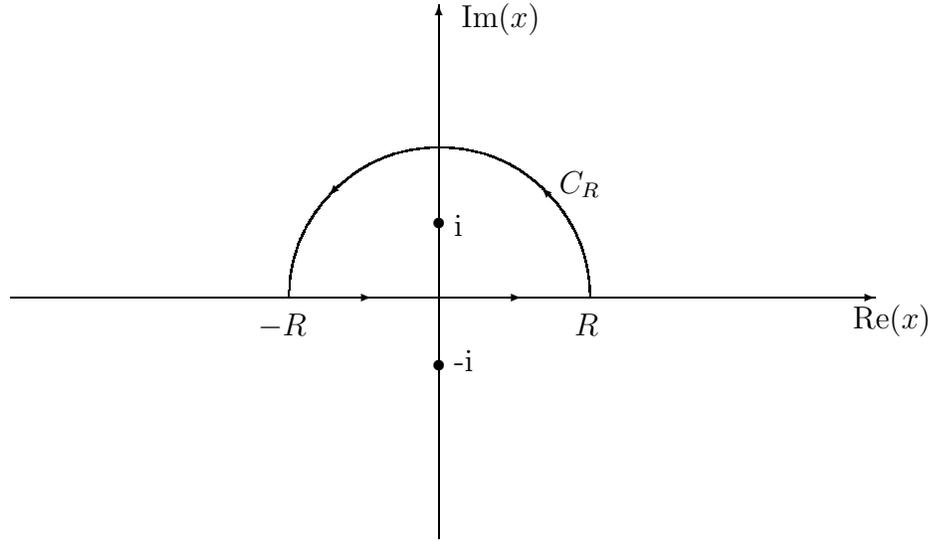
\setlength{\unitlength}{1mm}
\begin{picture}(120,70)
\put(20,39){\vector(1,0){115}}\put(132,35){Re($x$)}
\put(77,7){\vector(0,1){71}}\put(80,75){Im($x$)}
\put(77,30){\circle*{1.5}}\put(79,29.3){-i}
\put(77,49){\circle*{1.5}}\put(79,47.3){i}
\qbezier(97.0,39.0)(97.0,47.2843)(91.1421, 53.1421)
\qbezier(91.1421,53.1421)(85.2843,59.0)(77,59)
\qbezier(77.0,59.0)(68.7157,59.0)(62.8579,53.1421)
\qbezier(62.8579,53.1421)(57.0,47.2843)(57.0,39.0)
\put(53,34){$-R$}
\put(95,34){$R$}
\put(63,39){\vector(1,0){5}}
\put(83,39){\vector(1,0){5}}
\put(91.1421,53.1421){\vector(-1,1){.5}}
\put(62.8579,53.1421){\vector(-1,-1){.5}}
\put(93,53){$C_R$}
\end{picture}

By Cauchy's theorem, we have
\begin{equation}
\label{Cauchy}
\int_{-R}^{R}\!dx\; f(x) + \int_{C_R}\!dz\; f(z)=2\pi i\,\kappa_1
\end{equation}
where $\kappa_1$ is the residue of $f(z)$ at the (simple) pole $z=i$,
which, of course, can be calculated easily by
\begin{equation}
\kappa_1=(z-i)f(z)|_{z=i}=\left.\frac{1}{z+i}\right|_{z=i}=\frac{1}{2i}
\end{equation}

Therefore, for $R>1$, we have from (\ref{Cauchy})
\begin{equation}
\int_{-R}^{R}\frac{dx}{x^2+1}=\pi-\int_{C_R}\frac{dz}{z^2+1}
\end{equation}

Now, $|z|=R$ when $z$ is on the semi-circle $C_R$, so that
\begin{equation}
|z^2+1|\geq |z|^2-1 = R^2-1
\end{equation}
and therefore
\begin{equation}
\int_{C_R}\frac{dz}{z^2+1}\leq\int_{C_R}\frac{|dz|}{R^2-1}=\frac{\pi R}{R^2-1}\rightarrow 0 \:\:\: \mbox{for} \:\:\:R\rightarrow \infty
\end{equation}
Then, taking the limit,
\begin{equation}
\lim_{R\rightarrow \infty}\int_{-R}^{R}\frac{dx}{x^2+1}=\int_{-\infty}^{+\infty}\frac{dx}{x^2+1}=\pi
\end{equation}
or
\begin{equation}
I_1=\frac{1}{2}\int_{-\infty}^{+\infty}\frac{dx}{x^2+1}=\frac{\pi}{2}
\end{equation}

\section{Negative Dimension Integration Method}

On the other hand, let us take the Gaussian generating funtional of
the negative dimensional integration, namely,
\begin{equation}
\label{generating}
{\cal I}=\int_{-\infty}^{+\infty}\!dx\;{\rm e}^{-\alpha(x^2+1)}
\end{equation}
where we note that the argument in the exponential is the denominator
of the integrand in (\ref{I1}) multiplied by a real, positive
parameter $\alpha$ which is chosen as a converging factor for the
integral. Of course, this is the standard Gaussian integral, whose result is
\begin{equation}
\label{gaussian1}
{\cal I}={\rm
e}^{-\alpha}\pi^{\frac{1}{2}}\alpha^{-\frac{1}{2}}=\pi^{\frac{1}{2}}\sum_{j=0}^{\infty}(-)^j\,\frac{(\alpha)^{j-\frac{1}{2}}}{j!}
\end{equation}

If we project out in power series the integrand of (\ref{generating})
before the integration is done, we have
\begin{equation}
\label{sum}
{\cal I}=\sum_{n=0}^{\infty} (-)^n\frac{\alpha^n}{n!}\,I_{{\rm NDIM}}(n)
\end{equation}
where the negative dimensional integral $I_{{\rm NDIM}}(n)$ is given by
\begin{equation}
\label{NDIM}
I_{{\rm NDIM}}(n)\equiv \int_{-\infty}^{+\infty}\! dx\,(x^2+1)^n
\end{equation}

Note that for $n=-1$ this is exactly the integral (\ref{I1}) we want
to evaluate. Observe, however, that the negative dimensional integral
is more general than the one we have in (\ref{I1}) in the sense that
the exponent $n$ is not fixed. Note however that (\ref{NDIM}) is only
defined for positive $n$, so that in order to get the result for
(\ref{I1}) we need to make an analytic continuation to negative values
of $n$, a process whereby the integral gets defined into positive
dimensionality in (\ref{NDIM}).

Comparing the two series expansion (\ref{gaussian1}) with (\ref{sum}),
we observe that in order to both series be equivalent, we must have
$n=j-\frac{1}{2}$, so that
\begin{eqnarray}
I_{{\rm NDIM}}(n)&\equiv &\int_{-\infty}^{+\infty}\! dx\,(x^2+1)^n=(-)^{-n}\,n!\,\pi^{\frac{1}{2}}\frac{(-)^{n+\frac{1}{2}}}{(n+\frac{1}{2})!}\nonumber \\
&=&(-\pi)^{\frac{1}{2}}\frac{\Gamma(n+1)}{\Gamma(n+\frac{1}{2}+1)}\nonumber\\
&=&\frac{(-\pi)^{\frac{1}{2}}}{(n+1)_{\frac{1}{2}}}
\end{eqnarray}

Now, analytic continuing ({\rm AC}) the exponent $n$ into negative values using
the property (\ref{Poch}), we have
\begin{equation}
I_{{\rm NDIM}}^{({\rm AC})}(n)=\int_{-\infty}^{+\infty}\!\frac{dx}{(x^2+1)^{-n}}=\pi^{\frac{1}{2}}\,(-n)_{-\frac{1}{2}}
= \pi^{\frac{1}{2}}\frac{\Gamma(-n-\frac{1}{2})}{\Gamma(-n)}
\end{equation}

Now, substituting $n=-1,-2,-3,...$, we get
\begin{eqnarray}
I_{{\rm NDIM}}^{({\rm AC})}(-1)&=&\int_{-\infty}^{+\infty}\!\frac{dx}{x^2+1}=\pi\,,\\
I_{{\rm NDIM}}^{({\rm AC})}(-2)&=&\int_{-\infty}^{+\infty}\!\frac{dx}{(x^2+1)^2}=\frac{\pi}{2}\\
I_{{\rm NDIM}}^{({\rm AC})}(-3)&=&\int_{-\infty}^{+\infty}\!\frac{dx}{(x^2+1)^3}=\frac{3\pi}{8} \\
\vdots\qquad &=& \qquad \qquad \vdots \qquad = \:\:\vdots \nonumber
\end{eqnarray}
and so on and so forth. Of course, for the original integrals, which
range from $[0,\infty)$, the corresponding values are half of the
values quoted above.

A more interesting case is the evaluation of the other integral, namely,
\begin{eqnarray}
I_2&=&\int_0^{\infty}\!\frac{dx \;x^2}{(x^2+1)(x^2+4)}
=\frac{1}{2}\int_{-\infty}^{+\infty}\!\frac{dx\;x^2}{(x^2+1)(x^2+4)}
\end{eqnarray}

Evaluation of this integral by the residue technique is similar to the
previous one where now we have two poles in the upper hemisphere,
$z=i$ and $z=2i$, so that the contour $C_R$ is such that its radius
must be $|z|=R, R>2$, before taking the limit $R\rightarrow \infty$.
The result, after summing the residues of both poles is
\begin{equation}
I_2 = \frac{1}{2}\;2\pi i\left\{-\frac{1}{6i}+\frac{1}{3i}\right\} = \frac{1}{2}\left\{-\frac{\pi}{3}+\frac{2\pi}{3}\right\}=\frac{\pi}{6}.
\end{equation}

We have written down the explicit contributions of each residue, the
first one corresponding to the residue at the simple pole $z=i$ and
the second one to the residue at $z=2i$, to show that these residues
correlate with each of the {\em ``basis''} (linearly independent)
solutions (with a word borrowed from the language of basis vectors in a
vector space) in {\rm NDIM}.

The Gaussian generating functional of the negative dimensional
integral to this case is
\begin{eqnarray}
{\cal J}&\equiv&\int_{-\infty}^{+\infty}\!\!\!dx\;{\rm e}^{-\alpha x^2-\beta(x^2+1)-\gamma(x^2+4)}
= {\rm e}^{-\beta-4\gamma}\int_{-\infty}^{+\infty}\!\!\!dx\;{\rm e}^{-(\alpha+\beta+\gamma)x^2}\nonumber \\
&=&{\rm e}^{-\beta-4\gamma}\frac{\pi^{\frac{1}{2}}}{(\alpha+\beta+\gamma)^{\frac{1}{2}}}\nonumber\\
&=&\frac{\pi^{\frac{1}{2}}}{(\alpha+\beta+\gamma)^{\frac{1}{2}}}\sum_{r,s=0}^{\infty}(-)^{r+s}\frac{\beta^r}{r!}4^s\frac{\gamma^s}{s!}\nonumber\\
&=&\pi^{\frac{1}{2}}\Gamma(\frac{1}{2})\sum_{r,s,a,b,c}^{a+b+c=-\frac{1}{2}} (-)^{r+s}4^s \frac{\alpha^a\beta^{b+r}\gamma^{c+s}}{a!b!c!r!s!}\label{a}
\end{eqnarray}
where in the last line of the above we have employed the standard multinomial expansion for $(\alpha+\beta+\gamma)^{-\frac{1}{2}}$.

On the other hand, direct expansion in power series of the integrand yields
\begin{eqnarray}
\label{b}
{\cal J}&=&\sum_{j,l,m=0}^{\infty}(-)^{j+l+m}\frac{\alpha^j\beta^l\gamma^m}{j!\,l!\,m!}\int_{-\infty}^{+\infty}\!\!\!dx\,(x^2)^j(x^2+1)^l(x^2+4)^m\nonumber\\
&=&\sum_{j,l,m=0}^{\infty}(-)^{j+l+m}\frac{\alpha^j\beta^l\gamma^m}{j!\,l!\,m!}I_{{\rm NDIM}}(j,l,m)
\end{eqnarray}

Comparing (\ref{a}) with (\ref{b}) we see that we must have
\begin{eqnarray}
j&=&a\nonumber \\
l&=&b+r\nonumber\\
m&=&c+s\nonumber\\
a+b+c&=&-\frac{1}{2}
\end{eqnarray}

Also, when we compare the two series, the one in (\ref{a}) has five
summation indices with one constraint, whereas the other one (\ref{b})
has three summation indices. Then our result for $I_{{\rm NDIM}}(,j,l,m)
$ will be given in terms of $5-3-1=1$, that is, a single summation
index. However, since we have five indices in one and three in the
other, with one constraint equation among them, one can in principle
have the following combinatorial possibilities:
\begin{equation}
_5C_4=\frac{5!}{4!\,1!}=5
\end{equation}
for the remaining series index.

Letting $r$ be the summation index in the result, we have then the following conditions:
\begin{eqnarray}
a&=&j\nonumber\\
b&=&l-r\nonumber\\
c&=&-j-l-\frac{1}{2}+r\nonumber\\
s&=&j+l+m+\frac{1}{2}-r
\end{eqnarray}
so that the negative dimensional integral will be given by
\begin{eqnarray}
I_{{\rm NDIM}}^{(r)}(j,l,m)&=&(-)^{-j-l-m}j!\,l!\,m!\,\pi^{\frac{1}{2}}
\Gamma(\frac{1}{2})\nonumber\\
&& \times \sum_{r=0}^{\infty}\frac{(-)^{j+l+m+\frac{1}{2}}\,4^{j+l+m+\frac{1}{2}-r}}{j!\,(l-r)!\,(-j-l-\frac{1}{2}+r)!\,r!\,(j+l+m+\frac{1}{2}-r)!}\nonumber\\
&=&(-\pi)^{\frac{1}{2}}4^{j+l+m+\frac{1}{2}}\frac{\Gamma(\frac{1}{2})\Gamma(1+m)}{\Gamma(1-j-l-\frac{1}{2})\Gamma(1+j+l+m+\frac{1}{2})}\nonumber\\
&&\times \sum_{r=0}^{\infty}\frac{1}{(1+l)_{-r}(1-j-l-\frac{1}{2})_r(1+j+l+m+\frac{1}{2})_{-r}4^rr!}\nonumber\\
&=&\frac{(-\pi)^{\frac{1}{2}}\,4^{j+l+m+\frac{1}{2}}}{(\frac{1}{2})_{-j-l}\;(1+m)_{j+l+\frac{1}{2}}}\:\sum_{r=0}^{\infty}\frac{(-l)_r\;(-j-l-m-\frac{1}{2})_r}{(1-j-l-\frac{1}{2})_r}\frac{(\frac{1}{4})^r}{r!}\nonumber\\
&=&\frac{(-\pi)^{\frac{1}{2}}\,4^{j+l+m+\frac{1}{2}}}{(\frac{1}{2})_{-j-l}\;(1+m)_{j+l+\frac{1}{2}}}\: {}_2F_1\left(
\left.
\begin{array}{c}
-l\:,\: -j-l-m-\frac{1}{2} \\
1-j-l-\frac{1}{2}
\end{array}
\right |\frac{1}{4} 
\right)
\end{eqnarray}

Analogous calculations can be performed for $s,b,c$ indices, whereas
for the index $a$ there is no solution except the trivial one (This is
the case when the determinant of the system of linear equations
vanishes). For the remaining non-vanishing determinants, we have then
the solutions: For $s$ remaining summation index, we have the conditions
\begin{eqnarray}
a&=&j\nonumber\\
b&=&-j-m-\frac{1}{2}+s\nonumber\\
c&=&-m-s\nonumber\\
r&=&j+l+m+\frac{1}{2}-s
\end{eqnarray}
and the solution yields:
\begin{eqnarray}
I_{{\rm NDIM}}^{(s)}(j,l,m)&=&\frac{(-\pi)^{\frac{1}{2}}}{(\frac{1}{2})_{-j-m}\;(1+l)_{j+m+\frac{1}{2}}}\:\sum_{s=0}^{\infty}\frac{(-m)_s\;(-j-l-m-\frac{1}{2})_s}{(1-j-m-\frac{1}{2})_r}\frac{4^s}{s!}\nonumber\\
&=&\frac{(-\pi)^{\frac{1}{2}}}{(\frac{1}{2})_{-j-m}\;(1+l)_{j+m+\frac{1}{2}}}\:{}_2F_1\left(
\left.
\begin{array}{c}
-m\:,\: -j-l-m-\frac{1}{2} \\
1-j-m-\frac{1}{2}
\end{array}
\right|4 \right)
\end{eqnarray}

For $b$ summation, we have
\begin{eqnarray}
a&=&j\nonumber\\
r&=&l-b\nonumber\\
s&=&j+m+\frac{1}{2}+b\nonumber\\
c&=&-j-\frac{1}{2}-b
\end{eqnarray}
and the solution yields:
\begin{eqnarray}
I_{{\rm NDIM}}^{(b)}(j,l,m)&=&\frac{(-\pi)^{\frac{1}{2}}\:4^{j+m+\frac{1}{2}}}{(\frac{1}{2})_{-j}\;(1+m)_{j+\frac{1}{2}}}\:{}_2F_1\left(
\left.
\begin{array}{c}
j+\frac{1}{2}\:,\: -l \\ 1+j+m+\frac{1}{2}
\end{array}
\right|4 \right)
\end{eqnarray}

Finally, for $c$ summation we have
\begin{eqnarray}
a&=&j\nonumber\\
r&=&j+l+\frac{1}{2}+c\nonumber\\
s&=&m-c\nonumber\\
b&=&-j-\frac{1}{2}-c
\end{eqnarray}
and the solution yields:
\begin{eqnarray}
I_{{\rm NDIM}}^{(c)}(j,l,m)&=&\frac{(-\pi)^{\frac{1}{2}}\,4^m}{(\frac{1}{2})_{-j}\;(1+l)_{j+\frac{1}{2}}}\:{}_2F_1\left(
\left.
\begin{array}{c}
j+\frac{1}{2}\:,\: -m \\ 1+j+l+\frac{1}{2}
\end{array}
\right|\frac{1}{4} \right)
\end{eqnarray}

Now, observe that we have four non-vanishing solutions arising from
the solving of systems of linear equations, which here clearly comes
in pairs when we look at the argument of the hypergeometric functions
in the results. The solution for the integral is then the sum of the
pairs with the same argument (linear combination), namely,
\begin{eqnarray}
I_{{\rm NDIM}}(j,l,m)&=&A\:{}_2F_1(a,b;c|z^{-1})+B\:{}_2F_1(d,e;f|z^{-1})\\
               &=&A'\:{}_2F_1(a',b';c'|z)+B'\:{}_2F_1(d',e';f'|z)
\end{eqnarray}
where the coefficients $A$, $B$, $A'$ and $B'$ are given by
\begin{eqnarray}
A&=&\frac{(-\pi)^{\frac{1}{2}}\,4^{j+l+m+\frac{1}{2}}}{(\frac{1}{2})_{-j-l}\,(1+m)_{j+l+\frac{1}{2}}},\nonumber\\
B&=&\frac{(-\pi)^{\frac{1}{2}}\,\,4^m}{(\frac{1}{2})_{-j}\,(1+l)_{j+\frac{1}{2}}};\nonumber\\
A'&=&\frac{(-\pi)^{\frac{1}{2}}}{(\frac{1}{2})_{-j-m}\,(1+l)_{j+m+\frac{1}{2}}},\nonumber\\
B'&=&\frac{(-\pi)^{\frac{1}{2}}\:4^{j+m+\frac{1}{2}}}{(\frac{1}{2})_{-j}\,(1+m)_{j+\frac{1}{2}}}.
\end{eqnarray}
\noindent and the hypergeometric function parameters and variables are:

\vspace{.5cm}

\begin{center}
\begin{tabular}{|c|c|c||c|c|c|c|}\hline
$a$ & $b$ & $c$ & $d$ & $e$ & $f$ & $z^{-1}$ \\ \hline
$-l$ & $-j-l-m-\frac{1}{2}$ & $1-j-l-\frac{1}{2}$ & $-m$ &
$j+\frac{1}{2}$ & $1+j+l+\frac{1}{2}$ & $\frac{1}{4}$\\ \hline\hline
$a'$ & $b'$ & $c'$ & $d'$ & $e'$ & $f'$ & $z$ \\ \hline
$-m$ & $-j-l-m-\frac{1}{2}$ & $1-j-m-\frac{1}{2}$ & $-l$ &
$j+\frac{1}{2}$ & $1+j+m+\frac{1}{2}$ & $4$\\ \hline
\end{tabular}
\end{center}

\vspace{.5cm}

Before proceeding, let us demonstrate that the two sets of solutions,
namely the primed and unprimed ones are totally equivalent. To do
this, we employ the following analytic continuation property of
hypergeometric functions \cite{bateman}
\begin{eqnarray}
{}_2F_{1}(a,b;c|z)&=&\frac{z^{-a}}{(1-b)_a\;(c)_{-a}}\;{}_2F_{1}(a,1+a-c;1+a-b|z^{-1})\nonumber\\
&+&\frac{z^{-b}}{(1-a)_b\;(c)_{-b}}\;{}_2F_{1}(b,1+b-c;1+b-a|z^{-1}),\qquad |arg(-z)|<\pi
\end{eqnarray}
to one of the basis solutions, say of the unprimed set
\[
{}_2F_1(-l,\,-j-l-m-\textstyle\frac{1}{2};\,1-j-l-\textstyle\frac{1}{2}\,|\,\textstyle\frac{1}{4})
\]
to get the primed result. Of course, we could have used the
transformation property above to the other basis solution
${}_2F_1(-m,\,j+\frac{1}{2};\,1+j+l+\frac{1}{2}\;|\frac{1}{4})$, and
we would get the same primed result.  We need to apply only to one of
the basis solutions, since the transformation property above referred
to cannot produce neither new nor any more than two linearly
independent hypergeometric functions.

Now, we need to analytic continue the results to negative values of
exponents and positive dimension. Observe that the result for $I_{{\rm
NDIM}}$ contains two factors: One is the coefficients, given by ratios
of gamma functions and the other is the functional part, given by the
hypergeometric funtions. For the coefficients, which contain ratios of
gamma functions given in terms of Pocchhammers symbols, we employ
(\ref{Poch}), and for the functional part, just let the exponents go
to negative valued parameters. Note that we need to be aware of which
exponent should be continued to negative values \cite{tensorial}. Then, our final result
for the integral reads:
\begin{eqnarray}
I_{{\rm NDIM}}^{{\rm AC}}(j,l,m)&=&A^{{\rm AC}}\:{}_2F_1(a,b;c|z^{-1})+B^{{\rm AC}}\:{}_2F_1(d,e;f|z^{-1})\\
               &=&A'^{{\rm AC}}\:{}_2F_1(a',b';c'|z)+B'^{{\rm AC}}\:{}_2F_1(d',e';f'|z)
\end{eqnarray}
where 
\begin{eqnarray}
A^{{\rm AC}}&=&\pi^{\frac{1}{2}}\:4^{j+l+m+\frac{1}{2}}\;(\textstyle\frac{1}{2})_{j+l}\;(-m)_{-j-l-\frac{1}{2}},\nonumber\\
B^{{\rm AC}}&=&\pi^{\frac{1}{2}}\:4^m\;(\textstyle\frac{1}{2})_{j}\;(-l)_{-j-\frac{1}{2}};\nonumber\\
A'^{{\rm AC}}&=&\pi^{\frac{1}{2}}\:(\textstyle\frac{1}{2})_{j+m}\;(-l)_{-j-m-\frac{1}{2}},\nonumber\\
B'^{{\rm AC}}&=&\pi^{\frac{1}{2}}\:4^{j+m+\frac{1}{2}}\:(\textstyle\frac{1}{2})_{j}\;(-m)_{-j-\frac{1}{2}}.
\end{eqnarray}

One interesting thing about the NDIM techmology is that it allows us
to write the correct answer in as many equivalent ways as it is
possible to do. For the case in question, we have two equivalent
answers, namely, the unprimed and primed answers.

So, for particular values of the exponents, say, $j=1$ and $l=m=-1$,
which is the case for the $I_2$ integral mentioned in the
introduction, we have:
\begin{eqnarray}
I_{{\rm NDIM}}^{{\rm AC}}(1,-1,-1)&=&\pi^{\frac{1}{2}}\left\{4^{-\frac{1}{2}}\:(1)_{-\frac{1}{2}}\:{}_2F_1(1,\,\textstyle\frac{1}{2};\,\textstyle\frac{1}{2}|\frac{1}{4})+4^{-1}\:(\textstyle\frac{1}{2})_1\;(1)_{-\frac{3}{2}}\;{}_2F_1(1,\,{\small \frac{3}{2}};\,\frac{3}{2}|\frac{1}{4})\right\} \label{respa}\\
&=&\pi^{\frac{1}{2}}\left\{(1)_{-\frac{1}{2}}\:{}_2F_1(1,\,\textstyle\frac{1}{2};\,\textstyle\frac{1}{2}|4)+4^{\frac{1}{2}}\:(\textstyle\frac{1}{2})_1\;(1)_{-\frac{3}{2}}\;{}_2F_1(1,\,\frac{3}{2};\,\frac{3}{2}|4)\right\}. \label{respb}
\end{eqnarray}

Finally, using the fact that \cite{bateman}
\begin{equation}
{}_2F_1(a,b;b|z)=(1-z)^{-a}
\end{equation}
the two results (\ref{respa}) and (\ref{respb}) coalesce into one:
\begin{equation}
I_{{\rm NDIM}}^{{\rm AC}}(1,-1,-1)=-\frac{\pi}{3}+\frac{2\,\pi}{3}=\frac{\pi}{3}
\end{equation}
so that each of the basis solution corresponds exactly to the residue of the
poles at $z=i$ and $z=2i$. Finally,
\begin{equation}
I_2=\frac{1}{2}I_{{\rm NDIM}}^{{\rm AC}}(1,-1,-1)=\frac{\pi}{6}.
\end{equation}

Other particular cases, such as $j=m=0,\,l=-1$, $j=0,\,l=m=-1$ and
$j=0,\,l=-1,\,m=-2$ can be calculated from the general solution,
yielding,
\begin{eqnarray*}
I_1&=&\textstyle\frac{1}{2}\,I_{{\rm NDIM}}^{{\rm AC}}(0,-1,0)=\textstyle\frac{\pi}{2};\\
I_2^{(0,1,1)}&=&\textstyle\frac{1}{2}\,I_{{\rm NDIM}}^{{\rm AC}}(0,-1,-1)=\textstyle\frac{\pi}{12};\\
I_2^{(0,1,2)}&=&\textstyle\frac{1}{2}\,I_{{\rm NDIM}}^{{\rm AC}}(0,-1,-2)=\textstyle\frac{5\,\pi}{288}.\\
\end{eqnarray*}

Note that for all the cases where either $l=0$ or $m=0$ the general
solution is such that one of the terms in the overall result vanishes
because we have a term proportional to $\frac{1}{\Gamma(0)}=0$, and
the answer is then given by a single hypergeometric function.
 
\section{Conclusions}

Using the NDIM technique, we evaluated some sample real definite
integrals which may be calculated by the Cauchy residue theorem in the
complex plane. The alternative methodology here presented gives us the
bonus in that all the generic exponents of integrands can be
calculated at once, from where particular solutions can be
drawn. There is no difficulty in the performing of the integration
since the integration involved is of the Gaussian type and the
technique requires only series comparison term by term and the
solving of systems of algebraic linear equations resulting from such a
comparison. We showed that the results for different variables are
obtained simultaneously and they are equivalent to each
other. Moreover, for each set of basis solutions correspond the
residue of a given pole. The strength of this new technique also can be
envisaged in that simple, double or higher order poles can be
evaluated all at once.

\section{Acknowledgments}

The author gratefully acknowledges the kind hospitality of the
Department of Physics, North Carolina State University, and finantial
support from CAPES (Bras\'{\i}lia).

\end{document}